\documentclass[showpacs,aps]{revtex4-2}

\usepackage[dvips]{graphicx}
\usepackage{bm}
\usepackage{mathrsfs}
\usepackage{amsmath}
\usepackage{amsfonts}

\begin{document}
\date{\today} 
\title{Uncertainty Relation and Minimum Wave Packet on Circle}
\author{N. Ogawa\dag \: and S. Nagasawa\ddag 
\footnote{ogawanao@hus.ac.jp, nagasawa@hakodate-ct.ac.jp}}
\affiliation{\dag Hokkaido University of Science, 7-15 Maeda, Teine-ku, Sapporo 006-8585 Japan \\
\ddag National Institute of Technology, Hakodate College, 14-1 Tokura-cho, Hakodate 042-8501 Japan }

\begin{abstract}
We discuss on the uncertainty relation (UR) for a closed one dimensional system (circle).  
In such a system, we cannot use the angle  along the circle as a position variable.
Otherwise we meet difficulties about the definition of the average position and the standard deviation (SD), and Hermitian property of angular momentum.  From these reasons, we define the  position variable as Cartesian variable $(X,Y)$ that have the periodic property for angle $\phi$.  In the same way we define a SD by using that variables. Then we obtain two URs. 
We also discuss the minimum wave packet (MWP) on the circle. MWPs are expressed by von Mises distribution functions.
Next we construct total URs by combining two URs for $X$ and $Y$. 
Furthermore, we extend the variables to $(X_n, Y_n)$ \; with $n= 1,2,3,\cdots$ and we have infinite series of total URs.  
We consider the meaning of such extended URs.
\end{abstract}
\maketitle

\section{Introduction}

\subsection{Periodic function as  dynamical variable}

In usual quantum mechanics, the wave functions in space and in momentum space are related to each other
 by the Fourier integral. Also, the standard deviations (hereafter abbreviated as SD) of the wave functions on space $\sigma_x$ and 
 in momentum space $\sigma_p$ are related to each other
by the uncertainty relation (hereafter abbreviated as UR), which follows  the property of the Fourier integral \cite{Schiff}-\cite{Messiah}.
On a space with periodicity such as a circle,  any single-valued function becomes periodic,
meaning that  we cannot obtain the usual UR \cite{Judge} - \cite{Fujikawa}.
This is because the UR between position and momentum is a consequence of the Fourier integral in an infinite region, 
but not the Fourier series on a periodic space. Therefore it is not straightforward  to obtain the UR on such a space. 
 One such example is a rotating system with angle $\phi $ around the Z-axis and angular momentum $L_z$.
The definition of $\phi$ is
\begin{equation}
\phi= \arctan (\frac{y}{x}).
\end{equation}
Then $\phi$ is defined only modulo $2\pi$. 
We define $\phi$ to be continuous from $-\infty$ to $+\infty$.

This system may satisfy

\begin{equation}
\hat{L}_z =  -i\hbar \frac{\partial}{\partial \phi},  \; \;  [\phi, \hat{L}_z ] = i\hbar. \label{eq:commutator}
\end{equation}

Then, we usually expect the existence of the UR

\begin{equation}
\sigma_\phi  \; \; \sigma_{L_z} \ge \frac{\hbar}{2}.  \label{eq:UR1}
\end{equation}

For the variable $\phi$, our definition of SD  is usual one .

\begin{equation}
\sigma_\phi \equiv \sqrt{<(\phi-<\phi>)^2>} = \sqrt{<\phi^2> -<\phi>^2}. \label{eq:msd}
\end{equation}

However,  eq. (\ref{eq:UR1}) is not true from the following consideration.

When we consider the angular momentum eigenstate, we have $\sigma_{L_z}=0$. But we know

\begin{equation}
\sigma_\phi \lesssim \pi,
\end{equation}

which contradicts eq. (\ref{eq:UR1}) \cite{Judge}.
Therefore,  a careful treatment of the UR is required.\\

Furthermore, we also have a contradiction for (\ref{eq:commutator}).
We should notice that $L_z$ is Hermitian only in Hilbert space 
$$\{\psi \}_p,$$
where the suffix $p$ denotes ``periodic function with period $2\pi$''.
That is, $\psi$ should satisfy the periodic boundary condition (PBC) 
\begin{equation}
\psi(\phi) = \psi(\phi+2\pi).  \label{eq:PBC}
\end{equation}

let us consider the eigen states of $L_z$: bra $<m'|$  and  the ket $|m>$ and check the validity of  (\ref{eq:commutator}).

\begin{equation}
<m'|[\phi, L_z]|m> =  i\hbar \delta_{m',m}.
\end{equation}

If $L_z$ is Hermitian, this gives to 

\begin{equation}
(m-m') <m'|\phi|m>= i\hbar  \delta_{m',m},
\end{equation}

which contradicts when $m=m'$.
The reason is that  $L_z$ is not  Hermitian since $\phi$ is not periodic \cite{Louisell}, \cite{Carruthers}.
More explicitly saying, we cannot neglect the total derivative term since the existence of $\phi$, and we have

\begin{equation}
<m'|L_z \phi |m> \neq <L_z m'| \phi |m> \;\;\; (\phi \; \psi_m (\phi) \not \in \{\psi\}_p).
\end{equation}

Next we consider  the normalization condition, which is  given by

\begin{equation}
\int_\beta^{\beta+2\pi} |\psi(\phi)|^2 d\phi =1,
\end{equation}

where the parameter $\beta$ is the integration boundary (or window \cite{Opatrny} ). 
Usually this integration does not depend on the boundary choice $\beta$ because of the periodic boundary condition eq.(\ref{eq:PBC}). \\

Next we define the mean value of any function $f(\phi)$ as

\begin{equation}
<f>_\beta \equiv \int_\beta^{\beta+2\pi} f(\phi) |\psi(\phi)|^2 d\phi.  
\end{equation}

This definition of the mean value of the function $f$ should not depend on the integration boundary $\beta$. 
To show that explicitly,  it is necessary to require a function $f$ to satisfy
$$\frac{d}{d\beta} <f>_\beta =0.$$

This equation leads to
$$ f(\beta) = f(\beta+2\pi), $$
where we utilized eq.(\ref{eq:PBC}). Since the boundary $\beta$ is any real value, we can simply write

\begin{equation}
f(\phi) = f(\phi+2\pi).  \label{eq:PBC2}
\end{equation}

Therefore if we want to consider the mean value of $f(\phi) $, it should have periodicity eq.(\ref{eq:PBC2}).  However we meet some cases that the function $f(\phi)$ is not periodic such as 
$\phi$ and $\phi^2$ that appear in (4). 
Let us consider the simple example. When the amplitude of wave function $|\psi|$ is uniform, then we have

\begin{equation}
|\psi| = \frac{1}{\sqrt{2\pi}} \;  \; \Rightarrow  \; \; <\phi>_\beta = \int_\beta^{\beta+2\pi} \phi |\psi|^2 d\phi = \beta + \pi    \label{eq:specialcase} 
\end{equation}

It is strange that  the mean position $<\phi>$ depends on unphysical boundary $\beta$ and uniquely determined on circle,  even though every points on circle are physically the same.

 Furthermore, when the amplitude of the wave function $|\psi|$ is nonuniform, the situation is more complex. Using eq.(\ref{eq:PBC}), we obtain

\begin{eqnarray}
<\phi>_\beta &\equiv& \int_\beta^{\beta+2\pi} \phi |\psi|^2 d\phi = <\phi>_0 + 2\pi \int_0^\beta |\psi|^2 d\phi,\\
<\phi^2>_\beta &=&  \int_\beta^{\beta+2\pi} \phi^2 |\psi|^2 d\phi = <\phi^2>_0 + 4\pi^2 \int_0^\beta |\psi|^2 d\phi +4\pi \int_0^\beta \phi |\psi|^2 d\phi,\\
(\sigma_\phi^\beta)^2 \; \; &=& <\phi^2>_\beta -(<\phi>_\beta)^2 \nonumber\\
 &=&  (\sigma_\phi^0)^2 + 4\pi (\pi -<\phi>_0) \int_0^\beta |\psi|^2 d\phi + 4\pi \int_0^\beta \phi |\psi|^2 dx\phi -4\pi^2 (\int_0^\beta |\psi|^2 d\phi)^2.
\end{eqnarray}

We find that the SD also depends on boundary $\beta$.
We call this ``integration boundary problem'' hereafter.

As we have shown  in (\ref{eq:PBC2}), one method to avoid the boundary dependence is to use the periodic function \cite{Carruthers}, \cite{Marc}. Therefore we will try the periodic function  $\tilde{\phi}$ instead of $\phi$ such as

\begin{equation}
\tilde{\phi} = \phi \qquad  (-\pi \le \phi < \pi), \qquad \mbox{with periodicity} \qquad \tilde{\phi} (\phi) = \tilde{\phi} (\phi+2\pi).
\end{equation}

\begin{figure}[h]
\begin{center}
\includegraphics[width=6cm]{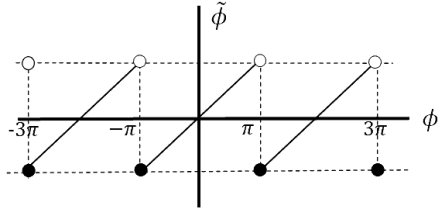}
\end{center}
\caption{Periodic function $\tilde{\phi}(\phi)$}
\end{figure}

And we consider

\begin{eqnarray}
<\tilde{\phi}>_\beta \; \;  &\equiv& \int_\beta^{\beta+2\pi} \tilde{\phi} |\psi|^2 d\phi =  \int_{-\pi}^{\pi} \phi |\psi|^2 d\phi =<\phi>,\\
<(\tilde{\phi})^2>_\beta &\equiv& \int_\beta^{\beta+2\pi} ( \tilde{\phi})^2  |\psi|^2 d\phi = 
\int_{-\pi}^{\pi} \phi^2 |\psi|^2 d\phi = <\phi^2>,\\
(\sigma_{\tilde{\phi}}^\beta)^2 \; \; &\equiv& <(\tilde{\phi})^2>_\beta-(<\tilde{\phi}>_\beta)^2 =  <\phi^2>- (<\phi>)^2 = (\sigma_\phi)^2.
\end{eqnarray}

Above three quantities do not depend on the boundary $\beta$.
However, in such a case, we have unwanted factor in commutation relation that is induced from the
discontinuity in $\tilde{\phi}$.

\begin{equation}
[\tilde{\phi}, L_z] = i\hbar (1-2\pi\sum_{-\infty}^{\infty}\delta[\phi -(2n+1)\pi]).
\end{equation}

We note that if we ignore this integration boundary problem and by using the selection of  the parameter $\beta=-\pi$, Kennard--Robertson (Hereafter abbreviated as KR)  method, \cite{Ken}, \cite{Rob}, leads us to the UR shown in Fujikawa's discussion \cite{Fujikawa},

$$ \sigma_\phi \;\; \sigma_{L_z} \ge \frac{\hbar}{2}(1-2\pi|\psi(\pi)|^2). $$

The evidence indicating that the SD depends on the boundary can be obtained from the r.h.s..\\

As we have seen above, if we utilize $\phi$ as position variable, we have ambiguities for mean position and SD.
For the Hermiticity of $L_z$ and to avoid the integration boundary dependence problem, our best selection of $\tilde{\phi}$
is the simple trigonometric function, \cite{Louisell}, \cite{Carruthers}, \cite{Marc}, \cite{Opatrny}.

\begin{eqnarray}
X &=& \cos \phi, \mbox{\; and } \\
Y &=& \sin \phi.
\end{eqnarray}

The geometrical meaning of this choice is as follows.
When we consider the uniform wave function case: eq. (\ref{eq:specialcase}), 
any points on circle have the same possibility to be the mean position. 
In other words, the mean value $<\phi>$ is indefinite though we have definite result eq. (\ref{eq:specialcase}) .

To solve this strange problem, it is suggestive to consider the center of mass (hereafter abbreviated as CM) of a curved wire as a mean value. If the wire is straight and has finite length, the CM is on the wire.  However, if the wire is curved, it is nor on the wire generally.
This fact suggests an idea that an averaged point might be outside the circle.
Then the mean position is expressed by the Cartesian coordinate. \cite{Opatrny}

\begin{equation}
( <X>, <Y> ).
\end{equation}

This idea is utilized in the next subsection.


\subsection{Position variable and expectation value}

   As we have seen in the previous  section, the mean value and  SD  of $\phi$ depends on a nonphysical parameter, i.e., the boundary $\beta$, 
and also $L_z$ is no longer Hermite which is very important requirement in quantum mechanics.
To avoid the difficulty, we use $\{X, Y \}$ as the basic position variable instead of $\phi$.

$$\vec{r} =  (X,Y) $$

 is Cartesian and the origin is taken as the center of the circle. 
These variables satisfies the periodicity $X(\phi) = X(\phi+2\pi), Y(\phi)= Y(\phi+2\pi)$.
The expectation value of $F(X,Y)$ is defined as 

\begin{equation}
<F>_\beta  \equiv \int_\beta^{\beta+2\pi} F(X,Y) \;  |\psi(\phi)|^2 d\phi.
\end{equation}

This definition of expectation value does not depend on boundary $\beta$. Therefore, hereafter we omit the suffix of $<\cdots>$.
We notice that 

$$X^2 + Y^2=1$$

holds, and we define ``the mean resultant length'' $R$, \cite{Opatrny}, \cite{Rade}  by

\begin{equation}
R \equiv \sqrt{<X>^2 + <Y>^2}  \le 1.
\end{equation}

The reason is simple.
From the fact $<X>^2 \le <X^2>$ and $<Y>^2 \le <Y^2>$,
we have
$$R=\sqrt{<X>^2 + <Y>^2} \le \sqrt{<X^2+Y^2>} =1.$$

Then we can define  $<\phi>$ when $R \neq 0$, such as  \cite{Opatrny}

\begin{eqnarray}
<X> &=& R \cos <\phi>, \label{eq:MRL1},\\
<Y> &=& R \sin <\phi>. \label{eq:MRL2}.
\end{eqnarray}

When the amplitude $|\psi|$ is uniform, we have $R=0$. Then $<\phi>$ is undefined as we expected, 
and the mean position is the origin of the circle.\\

 In the following, we discuss the URs and the minimum wave packet (hereafter abbreviated as MWP) of our periodic system with the generalization of  $X$ and $Y$ to $X_n$ and $Y_n$. In section II, we define the SD for $X_n$ and $Y_n$. Furthermore, utilizing the KR method, we introduce two kinds of  URs for any $n$. In section III, we discuss the form of MWPs.  In usual quantum mechanics, the MWP is expressed as a Gaussian function. 
However, in our circle case, our MWP is expressed by a von Mises distribution function which is usually referred to “Gaussian function on a circle”.  We have two URs and two MWPs. The two kinds of MWPs are similar, but one of them cannot act as other one's double.
 In section IV, we discuss the total UR by combining the two kinds of URs.  
We show that the total UR satisfies favorite properties as usual one, and we give three examples. 
 In section V, we give a discussion on the infinite URs due to the index $n$.


\section{Uncertainty relations for $X$ and $Y$}

We will extend the new position variable namely

\begin{equation}
X_n \equiv \cos (n\phi) ,  \; \; Y_n \equiv \sin (n\phi), \; \mbox{where}\; n=1,2,3, \cdots.  \label{eq:XY}
\end{equation}

This extension is natural because the periodicity still holds as

$$ X_n(\phi) = X_n(\phi + 2\pi), \;\;\; Y_n(\phi)= Y_n(\phi+2\pi). $$

Note that this extension is also discussed by Marc and Leblond in eqn. (2.11) - (2.17) \cite{Marc}.
We define the variance,

$$ \Delta X_n \equiv X_n-<X_n>,  \; \; \Delta Y_n \equiv Y_n -<Y_n>, \; \;  \Delta{\hat{L}_z}\equiv \hat{L}_z - <\hat{L}_z>.$$

Then, we can define two kinds of SDs

\begin{eqnarray}
&\sigma_{Xn}^2& \equiv~ <(\Delta X_n)^2>= <X_n^2>-<X_n>^2,\\
&& \nonumber\\
&\sigma_{Yn}^2& \equiv~ <(\Delta Y_n)^2>= <Y_n^2>-<Y_n>^2.
\end{eqnarray}

Our purpose is to construct an UR using $X_n$ and $Y_n$.
First, we express the commutation relations.

\begin{eqnarray}
&& [~X_n , ~ \hat{L}_z~ ] = - i n \hbar \; Y_n,  \label{eq:Xcom}\\
&& [~~Y_n  , ~ \hat{L}_z~ ] = + i n \hbar \; X_n. \label{eq:Ycom}
\end{eqnarray}

Then, we follow the KR method \cite{Ken}, \cite{Rob}.

We consider the quantity with a real parameter $\lambda$,

\begin{eqnarray}
&& ||(\Delta X_n -i\lambda \Delta \hat{L}_z)|\psi>||^2 \nonumber\\
&& = \int [ \{ (\Delta X_n-i\lambda \Delta \hat{L}_z) \psi \}^*  (\Delta X_n-i\lambda \Delta \hat{L}_z ) \psi ] dx \nonumber\\
&& = <\psi |(\Delta X_n +i\lambda \Delta \hat{L}_z)(\Delta X_n -i\lambda \Delta \hat{L}_z)|\psi> \ge 0.
\end{eqnarray}

The final form can be calculated as

\begin{eqnarray}
&& <\psi |(\Delta X_n +i\lambda \Delta \hat{L}_z)(\Delta X_n -i\lambda \Delta \hat{L}_z)|\psi>\nonumber\\
&=&  <\psi |(\Delta X_n)^2|\psi>+ \lambda^2  <\psi| ( \Delta \hat{L}_z)^2|\psi> -i \lambda  <\psi|[\Delta X_n, \Delta \hat{L}_z]|\psi>.
\end{eqnarray}

By using the relation

$$[\Delta X_n, \Delta \hat{L}_z]= [X_n, \hat{L}_z]= - i n \hbar Y_n, $$
we obtain the following  inequality for any real value of $\lambda$

$$ <\psi |(\Delta X_n)^2|\psi>+ \lambda^2  <\psi| ( \Delta \hat{L}_z)^2|\psi> -\lambda n \hbar <\psi|Y_n|\psi> \: \ge 0.$$

For this inequality to hold for any real value of $\lambda$, the discriminant should be less than or equal to 0.

$$ D = (n \hbar)^2  <\psi|Y_n|\psi>^2 - 4  <\psi| ( \Delta \hat{L}_z)^2|\psi>  <\psi |(\Delta X_n)^2|\psi> \le 0$$

This leads to the UR

\begin{equation}
 <\psi| ( \Delta \hat{L}_z)^2|\psi>  <\psi |(\Delta X_n)^2|\psi>\;  \ge \; \frac{(n \hbar)^2}{4}   <\psi|Y_n|\psi>^2,
\end{equation}

or, simply,

\begin{equation}
\sigma_{Xn}  \cdot  \sigma_{L_z}  \;  \ge \;  \frac{n \hbar}{2}  |<Y_n >|. \label{eq:ucr1}
\end{equation}

For $Y$, a similar treatment gives

\begin{equation}
\sigma_{Yn}  \cdot  \sigma_{L_z}  \;  \ge \;  \frac{n \hbar}{2} |<X_n >|.\label{eq:ucr2}
\end{equation}

Both eqs.  (\ref{eq:ucr1}) and  (\ref{eq:ucr2}) are the important results in this section.


\section{Minimum Wave Packet}
Let us  consider the Schwarz inequality to construct the UR \cite{Schiff}

\begin{equation}
|\int  f^* (x) g(x) dx|^2   \le   \int |f(x)|^2 dx  \cdot  \int |g(x)|^2 dx,
\end{equation}
where the equality holds when $f=\gamma g$ with a constant $\gamma$.\\

Furthermore, supposing $f = (X_n-<X_n>)\psi ~\mbox{and}~g=(\hat{L}_z-<\hat{L}_z>)\psi$, and changing $x$ by $\phi$, we have
\begin{equation}
|\int \psi^* (X_n-<X_n>)(\hat{L}_z-<\hat{L}_z>)\psi d\phi|^2 \le \int \psi^* (X_n-<X_n>)^2 \psi d\phi \cdot  \int \psi^* (\hat{L}_z-<\hat{L}_z>)^2 \psi d\phi.
\end{equation}

This shows that
\begin{eqnarray}
&& \frac{1}{4} |\int \psi^* (\{ \Delta X_n, \Delta \hat{L}_z \} + [ \Delta X_n, \Delta \hat{L}_z]) \psi d\phi|^2 \nonumber\\
&& \;  \;  \;   \le \; \; < (\Delta X_n)^2>  \;  < (\Delta \hat{L}_z)^2> = \sigma_{Xn}^2 \cdot \sigma_{L_z}^2.
\end{eqnarray}

Then we obtain
\begin{equation}
\frac{1}{4} |<\{ \Delta X_n, \Delta \hat{L}_z \}> -i n \hbar <Y_n>|^2 \; \;  \le \; \; \sigma_{Xn}^2 \cdot \sigma_{L_z}^2.
\end{equation}

The equality holds when

\begin{equation}
 (X_n-<X_n>)\psi_n = \gamma (\hat{L}_z-<\hat{L}_z>)\psi_n. \label{eq:req2}
\end{equation}

If  $\gamma$ is purely imaginary

$$\gamma = i a$$
with unknown real parameter $a$.  Then we can easily prove
\begin{equation}
<\{ \Delta X_n, \Delta \hat{L}_z \}>=0, \label{eq:req1}
\end{equation}

and we have 
\begin{equation}
\sigma_{Xn} \cdot \sigma_{L_z}  = \frac{n \hbar}{2}|<Y_n>|,
\end{equation}
which is the same as the minimum case of eqn. (\ref{eq:ucr1}).

Now we have
\begin{equation}
 (X_n-<X_n>)\psi_n = i a (\hat{L}_z-<\hat{L}_z>)\psi_n  \label{eq:req3}
\end{equation}
for the MWP.  Note that MWP for such a periodic system is ever discussed by Carruthers \cite{Carruthers}.
\\

This equation is transcribed to
\begin{eqnarray}
&& [a\hbar\frac{d}{d\phi}-\cos(n\phi)] \psi_n = \lambda_n \psi_n, \\
&& \lambda_n \equiv ia<\hat{L}_z>-<X_n>.
\end{eqnarray}

The integration can be performed simply to obtain

\begin{equation}
\psi_n = N_n \exp[\frac{1}{n a \hbar}\sin(n\phi)+i \frac{<\hat{L}_z>}{\hbar}\phi - \frac{<X_n>}{a\hbar} \phi]. \label{eq:wavepacket}
\end{equation}

 This function satisfies the periodic boundary condition only when 

\begin{equation}
<X_n>=0, \;\; <\hat{L}_z>=  m \hbar, \; m=0, \pm 1, \pm2, \cdots. \label{eq:auxiliary_condition}
\end{equation}

To check if the condition eq.(\ref{eq:auxiliary_condition}) is satisfied or not, we substitute the above conditions  into 
 eq.(\ref{eq:wavepacket}) and calculate $<\hat{L}_z>$ and $<X_n>$.

First, we consider the normalization

\begin{eqnarray}
 1 &=& \int_0^L |\psi_n (x)|^2 dx = |N_n|^2 \int_0^{2\pi} \exp[\frac{2}{n a \hbar}\sin(n\phi)] d\phi \nonumber\\
&= &|N_n|^2\int_0^{2\pi} \exp (\alpha_n \sin \theta) d\theta = 2\pi |N_n|^2  I_0 (\alpha_n),
\end{eqnarray}
where
$$\theta = n\phi, \;\;\; \alpha_n =\frac{2}{n a\hbar}. $$

Both $a$ and $\alpha_n$ take real values.  $I_0$  is the 0th deformed Bessel function of the first kind and is an even function. (see Appendix) Therefore, we take

\begin{equation}
N_n=  \frac{1}{\sqrt{2\pi I_0 (\alpha_n)}}.
\end{equation}

Then we have the MWP for $X_n$ with quantum number $(n,m)$  as

\begin{equation}
\psi_{X}^{(n,m)} = \frac{1}{\sqrt{2\pi I_0(\alpha_n)}}  \exp[\frac{\alpha_n}{2}\sin(n \phi)+i m\phi]. \; \; n=1,2,3,\cdots, \; m=0, \pm 1, \pm 2, \cdots \label{eq:wavepacket2}
\end{equation}
This function is called the von Mises distribution function. 
The form of $|\psi_{X}^{(n,m)}|$ is given in FIG. 2 for $n=1$ and $3$.
\\

The final task is to show explicitly that $<\hat{L}_z> = m \hbar$  and  $<X_n>=0$ by using eq. (\ref{eq:wavepacket2}).

First,
\begin{equation}
<\hat{L}_z>_X^{(n,m)} = \frac{m \hbar}{2\pi I_0(\alpha_n)} \int_0^{2\pi} \exp[\alpha_n \sin(n\phi)] d\phi =   m \hbar.
\end{equation}

This follows the fact that
$$I_0(\alpha) = \frac{1}{2\pi} \int_0^{2\pi} \exp[\alpha \sin \theta] d\theta, \; \;   \int_0^{2\pi} \cos \theta \;  \exp[\alpha \sin \theta]  \; d\theta =0.$$

Second, we also obtain
\begin{equation}
<X_n>_X^{(n,m)} =  \frac{1}{2\pi I_0(\alpha_n)} \int_0^{2\pi} \exp[\alpha_n \sin(n\phi)]  \cos(n\phi) d\phi =0. \label{eq:<X>}
\end{equation}

Thus, the condition  eq.(\ref{eq:auxiliary_condition}) is consistent with the form of MWP  eq.(\ref{eq:wavepacket}).
Note that $<Y_n>$ does not vanish, in contrast to $<X_n>$,

\begin{equation}
<Y_n>_X^{(n,m)} =  \frac{1}{2\pi I_0(\alpha_n)} \int_0^{2\pi} \exp[\alpha_n \sin(n\phi)]  \sin(n \phi) d\phi  = \frac{I_1(\alpha_n)}{I_0(\alpha_n)}.\label{eq:<Y>}
\end{equation}

\begin{figure}[h]
\begin{center}
\includegraphics[width=11cm]{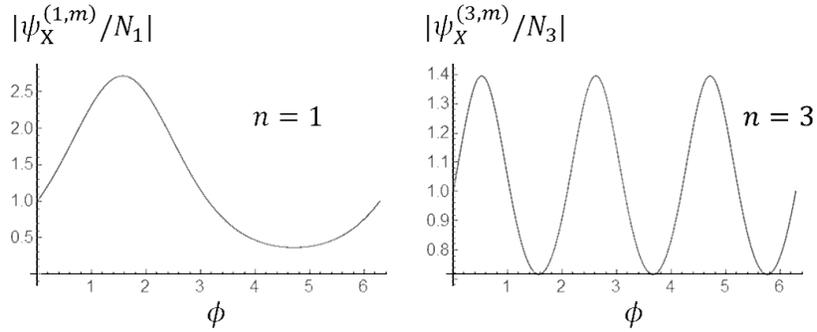}
\end{center}
\caption{MWPs of $n=1$ and $3$. $|\psi_X^{(n,m)} /N_n|$ is shown as a function of $\phi$, and we set $a \hbar=1$. 
It can be seen that $|\psi_X^{(n,m)}| $ has $n$ peaks.}
\end{figure}

Furthermore, we calculate $\sigma_{Xn}^2$ as

\begin{eqnarray}
\sigma_{Xn}^2 &=& <X_n^2>_X^{(n,m)} -<X_n>_X^{(n,m) 2} =  \frac{1}{2\pi I_0(\alpha_n)} \int_0^{2\pi}  \cos^2(n\phi) \exp[\alpha_n  \sin(n\phi)] d\phi \nonumber\\
&=& 1- \frac{1}{I_0(\alpha_n)}\frac{d^2 I_0(\alpha_n)}{d\alpha_n^2}= \frac{1}{\alpha_n}\frac{I_1}{I_0}=\frac{1}{\alpha_n}<Y_n>_X^{(n,m)},
\end{eqnarray}
where we utilized the recurring formula of the deformed Bessel function of the first kind (Appendix).

In total, we sum up the results on MWP for $X_n$.
\begin{equation}
\psi_{X}^{(n,m)} = \frac{1}{\sqrt{2\pi I_0(\alpha_n)}} \exp[\frac{\alpha_n}{2}\sin(n\phi)+i m\phi], \;\; \alpha_n = \frac{<Y_n>_{X}^{(n,m)}}{\sigma_{Xn}^2}, \; <X_n>_{X}^{(n,m)}=0. \label{eq:wavepacket3}
\end{equation}

For this MWP, we check the UR eq.(\ref{eq:ucr1}).

\begin{eqnarray}
\sigma_{L_z}^2 &=& <\hat{L}_z^2>_{X}^{(n,m)} -<\hat{L}_z>_{X}^{(n,m) 2} \nonumber\\
&=& -\frac{\hbar^2}{2\pi \;  I_0(\alpha_n)} \int_0^{2\pi} 
  e^{\frac{\alpha_n}{2} \sin n\phi -i m \phi} \frac{d^2}{d\phi^2} e^{\frac{\alpha_n}{2} \sin n\phi +i m\phi} d\phi  - ( m \hbar)^2 =  \alpha_n (\frac{ n \hbar}{2})^2 \frac{I_1(\alpha_n)}{I_0(\alpha_n)}.
\end{eqnarray}

Then, we have

\begin{equation}
\sigma_{Xn} \cdot \sigma_{L_z} |_{Xmwp} =  \frac{n \hbar}{2} |<Y_n>_{X}^{(n,m)}|. \label{eq:minimum_X}
\end{equation}

From this result,  the MWP  eq.(\ref{eq:wavepacket3}) is minimum for the UR  eq.(\ref{eq:ucr1})
but not for the UR  eq.(\ref{eq:ucr2}).  The reason is the following.

\begin{equation}
\sigma_{Y_n}^2 \: |_{Xmwp} =  <1-(X_n)^2>_{X}^{(n,m)} -<Y_n>_{X}^{(n,m) 2}  = 1-\frac{1}{\alpha_n} (\frac{I_1(\alpha_n)}{I_0(\alpha_n)}) - (\frac{I_1(\alpha_n)}{I_0(\alpha_n)})^2.
\end{equation}

Then, we have

\begin{equation}
\sigma_{Yn}^2 \cdot \sigma_{L_z}^2 \: |_{Xmwp} = (\frac{n \hbar}{2})^2 h(\alpha_n) \ge (\frac{n \hbar}{2})^2 <X_n>_{X}^{(n,m) 2} , \label{eq:UC}
\end{equation}
where

$$h(\alpha) \equiv \alpha (\frac{I_1(\alpha)}{I_0(\alpha)}) \{1-\frac{1}{\alpha} (\frac{I_1(\alpha)}{I_0(\alpha)}) - (\frac{I_1(\alpha)}{I_0(\alpha)})^2 \},$$

and the last inequality comes from the UR  eq.(\ref{eq:ucr2}) .

The form of function $h(\alpha)$ (even function) is shown in  FIG. 5 in Appendix.
As is shown clearly, $h(\alpha)$ is positive for any real number $\alpha$.

From
$$<X_n>_{X}^{(n,m)}=0,$$
the UR eq.(\ref{eq:UC}) holds.  Therefore, the equality holds in UR :eq. (\ref{eq:ucr2})  only when $\alpha=0 \; \mbox{or} \; \alpha \to \infty$ ( $h(\alpha) \to 0$).
\\

We end this section by putting all together  that we have discussed thus far.
First, we need new variables that satisfy a periodic boundary condition  to avoid the boundary dependence of the UR and to keep  $\hat{L}_z$  to be Hermite.  For this purpose, we use (\ref{eq:XY}) as new variables instead of $\phi$.

Second, when we utilize $X_n$ as a new dynamical variable, the commutation relations and URs are respectively  (\ref{eq:Xcom}) and (\ref{eq:ucr1}).
Third, the requirement of the minimum uncertainty for (\ref{eq:ucr1}) gives the MWP (\ref{eq:wavepacket3}).
\\

When we choose $Y_n$ as the dynamical variable, we can perform  similar calculations  (\ref{eq:Ycom}) and (\ref{eq:ucr2}).
The requirement of the minimum uncertainty for (\ref{eq:ucr2}) gives the MWP

\begin{equation}
\psi_Y^{(n,m)} = \frac{1}{\sqrt{2\pi I_0(\beta_n)}} \exp[-\frac{\beta_n}{2}\cos(n\phi) + i m\phi], 
\;\; \beta_n = \frac{<X_n>_Y^{(n,m)}}{\sigma_{Yn}^2}, <Y_n>_{Y}^{(n,m)}=0.  \label{eq: wavepacket4}
\end{equation}

Each variable $\{X_n,~Y_n\}$ has its own UR and  MWP.
Two wave packets $\{ \psi_X^{(n,m)}, \psi_Y^{(n,m)}\}$ are  essentially  the same except for a shift of the variable in the case of $m=0$.

\begin{equation}
\psi_Y^{(n,0)}(\phi) \sim \psi_X^{(n,0)} (\phi-\frac{\pi}{2n}) |_{(\alpha=\beta)}  \label{eq:phase}
\end{equation}


\section{Total Uncertainty and Examples}

In this section, we focus on the case of $n=1$.
From eqs. (\ref{eq:ucr1}) and  (\ref{eq:ucr2}),  we have
\begin{equation}
\sigma_{X}^2 \cdot \sigma_{L_z}^2 \ge (\frac{\hbar}{2})^2 <Y>^2,     \; \; \;      \sigma_{Y}^2 \cdot \sigma_{L_z}^2 \ge (\frac{\hbar}{2})^2 <X>^2. \label{eq:uncertain}
\end{equation}

In the previous sections we have defined two quantities.
The first one is the mean resultant length $R$  (hereafter abbreviated as MRL),
which is the length of the average vector on the unit circle, and the second one is the mean angle $<\phi>$ on circle as (\ref{eq:MRL1}) and (\ref{eq:MRL2}), \cite{Opatrny}.

Now we define the total SD $\tilde{\sigma}_R$ as follows

\begin{eqnarray}
\tilde{\sigma}_R^2 &\equiv&  \sigma_X^2 + \sigma_Y^2 = <X^2>-<X>^2 + <Y^2>-<Y>^2 \nonumber\\
&=& <X^2 + Y^2>-(<X>^2+<Y>^2) =  1-R^2.
\end{eqnarray}

By combining two inequalities in  eq. (\ref{eq:uncertain}), we obtain 

\begin{equation}
\tilde{\sigma}_R^2 \cdot  \sigma_{L_z}^2 \ge (\frac{\hbar}{2})^2 R^2. \label{eq:unctr}
\end{equation}

This is discussed by Marc and Leblond in eqs. (2.9) and (2.17) \cite{Marc}, and also by Holevo \cite{Holevo}.
The relation 

$$\tilde{\sigma}_R^2 = 1-R^2 \ge 0, \;\; R \ge 0$$

shows
\begin{equation}
 0 \le R \le 1, \; \; \mbox{and} \;  \; 0 \le \tilde{\sigma}_R \le 1.
\end{equation}

For $R=1$,  $\tilde{\sigma}_R =0$, there is no dispersion, and the position on circle is fixed by $<\phi>$. 
When $R<1$,  there is dispersion. When $R=0$,  the SD is maximum and  $\tilde{\sigma}_R=1$.
In such a case $<X>=<Y>=0$ and we identify the mean position at the origin of the circle.
\\

A more convenient definition of the total SD is
\begin{equation}
\sigma_R \equiv \frac{\sqrt{1-R^2}}{R} , \; \; \;  0 \le \sigma_R < \infty.
\end{equation}

Then, the UR follows eq.(\ref{eq:unctr}):

\begin{equation}
\sigma_R \cdot  \sigma_{L_z}  \ge \frac{\hbar}{2}. \label{eq:unctr2}
\end{equation}

To show the validity of eq.(\ref{eq:unctr2}), we consider three examples.
The first example is the super position of angular momentum eigen states \cite{Marc}. (Though the paper contains tiny mistake for this calculation)

\begin{equation}
\psi_1(\phi) =\frac{1}{\sqrt{4\pi}}(e^{ik\phi} + e^{im\phi}), \;\;\;  (k \neq m)
\end{equation}

with two different integers $k$ and $m$.
Then we have

\begin{equation}
<X>= \frac{1}{2}  \delta^1_{|k-m|}, \;\;\; <Y>=0.
\end{equation}

This gives 
$$R = \frac{1}{2} \; \delta^1_{|k-m|},$$
and

\begin{equation}
\sigma_R = \sqrt{3},
\end{equation}
for $|k-m|=1$. otherwise $\sigma_R = \infty$.
 Furthermore, 
\begin{equation}
\sigma_{L_z}=\frac{\hbar}{2} |k-m|.
\end{equation}
Then we obtain the result

\begin{equation}
\sigma_R \cdot \sigma_{L_z} = \frac{\sqrt{3}}{2} \hbar \ge \frac{1}{2}\hbar
\end{equation}
for  $|k-m|=1$

On the other hand, we obtain trivial relation
\begin{equation}
\sigma_R \cdot \sigma_{L_z} = \infty \ge \frac{1}{2}\hbar
\end{equation}
for $|k-m|\neq 1$ with $k \neq m$.

The second example is the $n$-th power of a trigonometrical function:

\begin{equation}
 \psi_2(\phi) = \sqrt{\frac{(2n)!!}{2\pi (2n-1)!!}} \;  \sin^n\frac{\phi}{2}, \qquad  n= 1,2,3, \cdots.
\end{equation}

Then, we obtain 
$$<X> = -\frac{n}{n+1}, \qquad <Y> = 0.$$

Furthermore,

$$ \sigma_R = \frac{\sqrt{2n+1}}{n}.$$
and
$$ \sigma_{L_z} = \frac{\hbar}{2} \frac{n}{\sqrt{2n-1}}.$$

In total, we obtain the UR, and the MWP is obtained in the limit $n \to \infty.$

\begin{equation}
\sigma_R \cdot \sigma_{L_z}= \frac{\hbar}{2} \sqrt{\frac{2n+1}{2n-1}} \ge \frac{\hbar}{2}.
\end{equation}

The third example is the von Mises distribution function, as MWP ($m=0$) for $X$ (\ref{eq:wavepacket2}).

\begin{equation}
 \psi_3(\phi) = \frac{1}{\sqrt{2\pi  I_0(\alpha)}} \exp [\frac{\alpha}{2} \sin \phi],
\end{equation}

From the previous calculation eqs.(\ref{eq:<X>}) and  (\ref{eq:<Y>}), we have

$$ <X>=0, \qquad  <Y> = \frac{I_1(\alpha)}{I_0(\alpha)}.$$

Then we obtain

$$R = \frac{I_1(\alpha)}{I_0(\alpha)},$$
and

$$ \sigma_{L_z} = \frac{\hbar}{2} \sqrt{\alpha  \frac{I_1(\alpha)}{I_0(\alpha)}}, \; \; \sigma_R = \sqrt{(\frac{I_0(\alpha)}{I_1(\alpha)})^2-1}.$$

Then we have UR 

\begin{equation}
\sigma_R \cdot \sigma_p = \frac{\hbar}{2} f(\alpha) \ge  \frac{\hbar}{2}, \; \; f(\alpha) \equiv \sqrt{\alpha (\frac{I_0(\alpha)}{I_1(\alpha)}-\frac{I_1(\alpha)}{I_0(\alpha)})}. \label{eq:uncertainty_trial}
\end{equation}

The form of $f(\alpha)$ is shown in FIG.3. The relation $f(\alpha) \ge 1$ is evident.
As we can see, the minimum uncertainty is obtained in the limit  $\alpha \to \infty$.
(Notice that MWP for X is NOT the ``minimum uncertainty'' for total UR.)

\begin{figure}[h]
\begin{center}
\includegraphics[width=7cm]{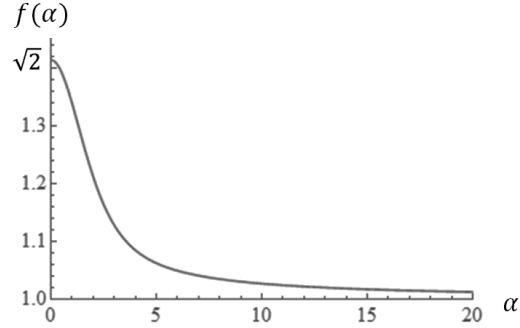}
\end{center}
\caption{Even function $f(x)$ which varies from $\sqrt{2}$ to 1 for positive $\alpha$.}
\end{figure}


\section{ Infinite Uncertainty Relations}
In the following, we will extend the uncertainty relation into the general form $(n \neq 1)$, and discuss the physical meaning.
We can simply extend the results obtained in the previous section IV for the case of $n \neq 1$. \cite{Marc}
We define $R_n$ by

\begin{equation}
R_n \equiv \sqrt{<X_n>^2 + <Y_n>^2}.
\end{equation}

and define SD by
\begin{equation}
\sigma_n \equiv \frac{1}{ n} \frac{\sqrt{1-R_n^2}}{R_n}.
\end{equation}

Then, we obtain infinite total URs by using eqs. (\ref{eq:ucr1}) and  (\ref{eq:ucr2}), 

\begin{equation}
\sigma_n \cdot \sigma_{L_z} \ge \frac{\hbar}{2}, \;\; n=1,2,3,\cdots.   \label{eq:unctr3}
\end{equation}

To consider the difference of UR due to the difference of $n$, we give one simple example. 
Let us consider the case as

\begin{equation}
\psi = \sqrt{\frac{1}{\pi}}\cos \phi.  \label{eq:2-fold}
 \end{equation}

In this case, we obtain $<X> = <Y>=0$ giving $R=0$.

$$ \sigma_1 = \infty, \qquad  \sigma_{L_z} = \hbar.$$

Therefore in the case of $n=1$, the UR gives the non sense result because UR for $n=1$ can be rewritten as

$$\sqrt{1-R^2} \; \sigma_{L_z} \ge \frac{\hbar}{2} R$$

and $R=0$ gives trivial result $\sigma_{L_z} \ge 0$.\\

On the other hand, in the case of $n=2$ for the same wave function,

$$<X_2> = \frac{1}{2}, \qquad  <Y_2> =0,$$

$$\sigma_2 = \frac{\sqrt{3}}{2}, \qquad  \sigma_{L_z} = \hbar.$$
 
This gives

$$ \sigma_2 \cdot \sigma_{L_z} = \frac{\sqrt{3}}{2} \hbar \ge \frac{\hbar}{2}.$$

Let us roughly explain the result obtained above.
In our case $|\psi|^2 \sim \cos^2 \phi$, the peaks are at $\phi = 0$ and $\pi$. 
Under this distribution, 
$X_1 = \cos \phi$ is $+1$ for $\phi = 0$, and $-1$ for $\phi = \pi$.
In total $<X_1> =0$. Next, $Y_1 = \sin \phi =0$ for both of $\phi = 0$, and $\pi$.
Therefore, $<Y_1> =0$ and thus we have $R_1=0, \;\; \sigma_1=\infty$.\\

However, for $n=2$ analysis, $X_2 = \cos 2\phi$ is $+1$ for both of $\phi=0$, and $\pi$.
That gives  $<X_2> >0$.  $Y_2= \sin 2\phi$ is $0$  for both of $\phi=0$, and $\pi$.
Thus we obtain $<Y_2> =0$. Accordingly we have $0<R_2<1$ and $\sigma_2$ is finite.
There might be some suitable $n$ for analysis  depending on the form of $\psi$.\\

To understand the situation more clearly, 
we consider the following n-fold $(n \ge 2)$ symmetric density function.
We will show  $R_1=0$ under such a symmetry.

First the $n$-fold symmetry shows

\begin{equation}
\rho(\phi) = \rho(\phi + \frac{2\pi}{n}), \;\;\; \rho(\phi) \equiv |\psi(\phi)|^2.
\end{equation}

Then the following quantity vanishes.

\begin{equation}
<X_1> +i <Y_1> = \int_0^{2\pi} e^{i\phi} \rho(\phi) d\phi=0,
\end{equation}

since
\begin{eqnarray}
\int_0^{2\pi} e^{i\phi} \rho(\phi) d\phi &=& \sum_{k=0}^{n-1} \int_{2\pi k/n} ^{2\pi (k+1)/n} e^{i\phi} \rho(\phi) d\phi. \\
&=& (\sum_{k=0}^{n-1} e^{i (2\pi/n)k})  \int_0^{2\pi/n} e^{i\theta} \rho(\theta) d\theta =0.
\end{eqnarray}

In such a case it is worth while to consider $R_n$.

\begin{eqnarray}
<X_n> +i <Y_n> &=& \int_0^{2\pi} e^{i n\phi} \rho(\phi) d\phi \nonumber \\
&=& \sum_{k=0}^{n-1} \int_{2\pi k/n} ^{2\pi (k+1)/n} e^{i n\phi} \rho(\phi) d\phi \nonumber \\
&=&n \int_0^{2\pi/n} e^{i n\theta} \rho(\theta) d\theta.
\end{eqnarray}

\begin{equation}
R_n = |<X_n> +i <Y_n>| =  n  |\int_0^{2\pi/n} e^{i n\theta} \rho(\theta) d\theta|.
\end{equation}

This relation has the following physical meaning.
When the density function has n-fold symmetry, the magnitude of $R_n$ shows the 
non-uniformity of density during period $\Delta \theta = \frac{2\pi}{n}$.
It is generally not equal to zero. Then n-th uncertainty relation makes a sense ($R_n \neq 0$).
Note that  there are the cases  $R_1=0$ without any  n-fold symmetry.
But they are not the case here we discuss.

At last we give one more example, which have 4-fold symmetric density function

\begin{equation}
\psi(\phi) = \frac{1}{\sqrt{\pi}}\cos 2\phi.
\end{equation}
The peak appears at $\phi=(0, \pi/2, \pi, 3\pi/2)$.

$$<\vec{r}_1> \equiv <X_1>\vec{i} + <Y_1>\vec{j} 
= \frac{1}{\pi}\int_0^{2\pi} \cos \phi \cos^2 (2\phi) d\phi \vec{i} 
                    +\frac{1}{\pi}\int_0^{2\pi} \sin \phi \cos^2(2\phi) d\phi \vec{j} =\vec{0},$$
 giving $R_1=0$.

 On the other hand,
 
 $$<\vec{r}_4> \equiv <X_4>\vec{i} + <Y_4>\vec{j} 
= \frac{1}{\pi}\int_0^{2\pi} \cos 4\phi \cos^2 (2\phi) d\phi \; \vec{i} 
                    +\frac{1}{\pi}\int_0^{2\pi} \sin 4\phi \cos^2(2\phi) d\phi \; \vec{j} =\frac{1}{2} \; \vec{i},$$
                    
giving $<X_4>=\frac{1}{2},<Y_4>=0$, and $R_4=\frac{1}{2}$.\\

Then we have 
\begin{equation}
\sigma_4= \frac{1}{4} \frac{\sqrt{1-R_4^2}}{R_4}=\frac{\sqrt{3}}{4}.
\end{equation}

Together with
$$\sigma_{L_z}=2\hbar,$$
we obtain

\begin{equation}
\sigma_4 \cdot \sigma_{L_z} = \frac{\sqrt{3}}{2}  \hbar \ge \frac{\hbar}{2}.
\end{equation}

 \section{Summary}

We will collect up what we have discussed above.\\

1. Meaning of average: We have redefined the method of taking the position average on a circle by using Cartesiian coordinate$(X, Y)$ with $X=\cos \phi,\; Y=\sin \phi$. The method is  similar to consider the CM of a curved wire: when the wire is curved, the CM is usually not on the wire \cite{Opatrny}. Then the integration boundary problem disappears , and Hermitian problem of angular momentum is solved \cite{Carruthers}.\\

2. Standard Deviation (SD) of position and Uncertainty Relation (UR): Any point on a circle is expressed by the Cartesian coordinate $(X,Y)$ with $X^2+Y^2=1$. Then, we can define the mean position by $(<X>,<Y>)$, and also the SD is defined for each axis as
$$\sigma_X = \sqrt{<X^2> -<X>^2}, \;\; \sigma_Y = \sqrt{<Y^2>-<Y>^2}.$$ We can also define the SD for the angular momentum operator as $\sigma_{L_z}$ and construct a pair of URs.

\begin{equation}
 \sigma_X  \cdot \sigma_{L_z} \ge \hbar/2, \; \; \sigma_Y \cdot \sigma_{L_z} \ge \hbar/2. \label{eq:2URs}
\end{equation}

3. Minimum Wave Packets (MWP): We have two kinds of uncertainty relations, one is for $\sigma_{L_z}$ and $\sigma_X$, and another is for $\sigma_{L_z}$ and $\sigma_Y$.  For each we can construct  the MWPs \cite{Carruthers}, \cite{Marc}. They are similar but one cannot be substituted by another. We also extend the system $(X,Y)$ to $(X_n, Y_n)$ with $X_n = \cos n\phi, \; Y_n = \sin n\phi$ and so we have infinite series of MWPs, that are shown explicitly in figure 2.\\

4. Total Uncertainty Relation (UR) :  We have shown that the total UR is simply given by  combining a set of URs (\ref{eq:2URs}). \cite{Carruthers}, \cite{Marc}.

\begin{equation}
 \sigma_R \cdot \sigma_{L_z} \ge \frac{\hbar}{2},  \;\;\;  \sigma_R = \frac{\sqrt{\sigma_X^2 + \sigma_Y^2}}{R} = \frac{\sqrt{1-R^2}}{R}, \;\;\; R \equiv \sqrt{<X>^2 + <Y>^2}.
\end{equation}

$R$ indicates the accuracy, $R=1$ means zero SD,  and $R=0$ means the maximum SD \cite{Opatrny}, \cite{Rade}. 
Three examples are shown to understand the total uncertainty relations.\\

5. Infinite total uncertainty relations: We have shown that infinite total URs exist due to $n$ index.

\begin{equation}
\sigma_n \cdot \sigma_{L_z} \ge \frac{\hbar}{2},  \;\;\;  \sigma_n = \frac{1}{n} \frac{\sqrt{1-R_n^2}}{R_n}, \;\;\; R_n \equiv \sqrt{<X_n>^2 + <Y_n>^2}.
\end{equation}

 These total URs are applicable when the density function $|\psi|^2$ has n-fold symmetry ($n \ge 2$). 
 Then we can show $R_1=0$ and $\sigma_1 = \infty$ and total UR makes no sense. 
 But  in such a case $R_n \neq 0$ generally, and $n-th$ UR makes a sense.

\section{Acknowledgement}

The authors greatly thanks to the referee who gave us very important informations and appropriate comments.
They also thanks to Prof. Yokoyama of Gakushuin University  for the discussion and thank Ms. Sasaki for her help in typing the manuscript.


\section{Appendix: Around the deformed Bessel function of  the first kind}

The 0th order deformed Bessel function of  the first kind is defined as

\begin{equation}
I_0(x) \equiv \frac{1}{2\pi} \int_0^{2\pi} \exp (x \sin \theta) d\theta = \frac{1}{2\pi} \int_0^{2\pi} \exp (x \cos \theta) d\theta.
\end{equation}
Note that this function is even function.

The 1st order deformed Bessel function is given by the derivative

$$ I_1(x) = \frac{d I_0(x)}{dx}= \frac{1}{2\pi} \int_0^{2\pi} \sin \theta  \; \exp (x \sin \theta) d\theta.$$

We also have the following recurring  formula, which we sometimes utilize:

$$x\frac{d I_n(x)}{dx} + n I_n(x) = x I_{n-1}.$$

The ratio of the two deformed Bessel function of the first kind,

$$\frac{I_1(x)}{I_0(x)},$$

 has the form shown in FIG.4, and its expansion around $x=0$ is

$$\frac{I_1(x)}{I_0(x)} \sim \frac{1}{2}x + {\cal O} (x^3).$$

and its asymptotic expansion at $x \to + \infty$ is

$$ \frac{I_1(x)}{I_0(x)} \sim 1-\frac{1}{2x} - \frac{1}{8 x^2} +  {\cal O}(x^{-3}).$$

\begin{figure}[h]
\begin{center}
\includegraphics[width=6cm]{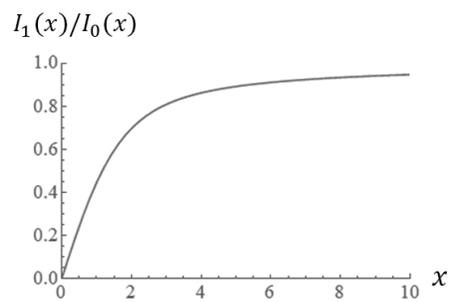}
\end{center}
\caption{Ratio of two modified Bessel functions of the first kind.}
\end{figure}

For the function
$$ h(x) \equiv x  (\frac{I_1(x)}{I_0(x)}) \{1-\frac{1}{x} (\frac{I_1(x)}{I_0(x)}) - (\frac{I_1(x)}{I_0(x)})^2 \},$$

we obtain its expansion around $x=0$ as
\begin{eqnarray}
h(x)&\sim&  \frac{x^2}{4} +  {\cal O}(x^{4}) ,
\end{eqnarray}

and its asymptotic expansion at $x \to + \infty$ is

\begin{eqnarray}
h(x) &\sim & \frac{1}{2x} +  {\cal O}(x^{-2}) .
\end{eqnarray}
 In total, for positive $x$, $h(x)$ has the form shown in FIG.5.

\begin{figure}[h]
\begin{center}
\includegraphics[width=7cm]{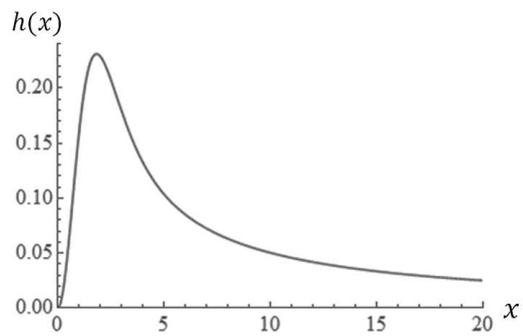}
\end{center}
\caption{ Form of function  $h(x)$.}
\end{figure}

\end{document}